\documentclass[11pt]{article}
\usepackage{times}
\usepackage{mathfont}
\usepackage{graphicx}
\usepackage{cite}
\usepackage{url}
\newtheorem{defn}{Definition}
\newtheorem{conjecture}{Conjecture}

\newtheorem{lemma}{Lemma}

\newtheorem{theorem}{Theorem}
\newtheorem{open}{Open Problem}
\newcommand{\qed}{~$\vrule width.15cm height.2cm depth0cm$ \medbreak}
\newenvironment{proof}{\noindent{\bf Proof: }}{\qed}

\def\log{\mathop{\rm log}}

\def\sin{\mathop{\rm sin}}

\DeclareSymbolFont{AMSb}{U}{msb}{m}{n}
\DeclareSymbolFontAlphabet{\Bbb}{AMSb}
\def\R{\ensuremath{\Bbb R}}
\def\Z{\ensuremath{\Bbb Z}}
\let\Real\R

\long\def\omit#1{}

\begin{document}

\title{Multivariate Regression Depth}

\author{
Marshall Bern
\thanks{Xerox Palo Alto Research Center,
3333 Coyote Hill Rd., Palo Alto, CA  94304.
\ bern@parc.xerox.com  }
\and
David Eppstein
\thanks{Dept.\ of Information and Computer Science,
U. of California, Irvine, CA 92697. 
\ eppstein@ics.uci.edu }
}

\date{}
\maketitle   
 
\begin{abstract}
The regression depth of a hyperplane with respect to
a set of $n$ points in $\Real^d$ is the minimum number of points
the hyperplane must pass through in a rotation to vertical.
We generalize hyperplane regression depth 
to $k$-flats for any $k$ between $0$ and $d-1$. The $k=0$ case gives the
classical notion of center points. We prove that for any $k$ and $d$,
deep $k$-flats exist, that is, for any set of $n$ points there always
exists a $k$-flat with depth at least a constant fraction of $n$. As a
consequence, we derive a linear-time $(1+\epsilon)$-approximation
algorithm for the deepest flat.
\end{abstract}

\section{Introduction}

Linear regression asks for 
an affine subspace (a {\it flat\/}) that fits a set of data points. 
The most familiar case assumes $d-1$ independent or {\em
explanatory} variables and one dependent or
{\em response} variable, and fits a hyperplane to explain the
dependent variable as a linear function of the independent variables. 
Quite often, however, there may be more than one dependent variable, and
the {\em multivariate regression} problem requires fitting a
lower-dimensional flat to the data points, perhaps even a succession of
flats of increasing dimensions. Multivariate least-squares regression is
easily solved by treating each dependent variable separately, but this is
not correct for other common forms of regression such as least absolute
deviation~\cite{Liu-Stat-92} or least median of
squares~\cite{Rou-JASA-84}.

Rousseeuw and Hubert~\cite{RouHub-JASA-99} introduced the notion
of {\it regression depth\/} as a robust criterion for linear
regression. The regression depth of a hyperplane $H$ fitting a set of $n$
points is  the minimum number of points whose removal makes $H$ into a
{\it nonfit\/}. A nonfit is a hyperplane that can be rotated to vertical 
(that is, parallel to the dependent variable's axis)
without passing through any points.
The intuition behind this definition 
is that a vertical hyperplane posits no relationship between the 
dependent and independent variables, and hence many points
should have to be invalidated in order to make a good regression hyperplane
combinatorially equivalent to a vertical hyperplane.
Since this definition does not make use of the size of the residuals,
but only uses their signs, it is robust in the face of skewed or
heteroskedastic (data-dependent) error models. Regression depth also has a
number of other nice properties including invariance under affine
transformations, and a connection to the classical notions of center
points and data depth. 

This paper generalizes regression depth
to the case of more than one dependent variable, that is, 
to fitting a $k$-flat to points in $\Real^d$.  
This generalization is not obvious:
for example, consider fitting a line to points in 
$\Real^3$.  Generic lines can be rotated wherever
one likes without passing through data points, so how are we
to distinguish one line from another?

We start by reviewing previous work (Section~\ref{prevsec})
and stating our basic definitions (Section~\ref{defsec}).
We then provide a lemma that may be of independent interest,
on finding a family of large subsets of any point set
such that the family has no hyperplane transversal
(Section~\ref{lemsec}).
We prove the existence of deep $k$-flats for any $k$ 
(Section~\ref{ksec}), and give tight bounds on depths of 
lines in $\R^3$ (Section~\ref{linesec}).  We conclude by
discussing related generalizations of Tverberg's theorem
(Section~\ref{tversec}),
describing possible connections between
$k$-flats and $(d-k-1)$-flats (Section~\ref{revsec}),
and outlining the algorithmic implications of our
existence proof (Section~\ref{algsec}).
Along with the results proven in each section,
we list open problems for further research.

\section{Previous Work}\label{prevsec}

Regression depth was introduced by Rousseeuw and
Hubert~\cite{RouHub-JASA-99} as a combinatorial measure of the quality of
fit of a regression hyperplane. An older notion, variously called
{\em data depth}, {\em location depth}, {\em halfspace depth}, or
{\em Tukey depth}, similarly measures the quality of fit of a
single-point estimator. It has long been known that there exists a point
of location depth at least $\lceil n/(d+1)\rceil$ (a
{\em center point}). Rousseeuw and Hubert provided a construction called
the {\em catline}~\cite{HubRou-JMA-98}
for computing a regression line
for a planar point set with depth at least
$\lceil n/3\rceil$, 
and conjectured~\cite{RouHub-DCG-99} that in higher dimensions as well
there should always exist a regression hyperplane of depth
$\lceil n/(d+1) \rceil$.
Steiger and Wenger~\cite{SteWen-CCCG-98}
proved that a deep regression hyperplane always exists,
but with a much smaller fraction than $\lceil 1/(d+1) \rceil$.
Amenta, Bern, Eppstein, and Teng~\cite{AmeBerEpp-DCG-?}
solved the conjecture using an argument based on 
Brouwer's fixed-point theorem and a close connection
between regression depth and center points.

On the algorithmic front,  
Rousseeuw and Struyf~\cite{RouStr-SC-98} gave algorithms for testing the
regression depth of a hyperplane.
Their time bounds are exponential in the dimension, unsurprising since
the problem is NP-complete for unbounded
dimension~\cite{AmeBerEpp-DCG-?}. For the planar case, Van
Kreveld, Mitchell, Rousseeuw, Sharir, Snoeyink, and Speckmann gave an
$O(n\log^2 n)$ algorithm for computing a deepest
line~\cite{KreMitRou-SCG-99}.  Langerman and Steiger
\cite{LanSte-SODA-00} later improved this to
an optimal $O(n\log n)$ time bound.

\section{Definitions}\label{defsec}

Although regression is naturally an affine rather than projective
concept, our constructions and definitions live most gracefully in
projective space. We view $d$-dimensional real projective space as a
renaming of objects in $(d+1)$-dimensional affine space.
(Affine space is the familiar Euclidean space, only we
have not specified a distance metric.)
A $k$-flat, for $1 \leq k \leq d$,
through the origin of $(d+1)$-dimensional affine
space is a {\it projective $(k-1)$-flat\/}. 
In particular a line through the origin is a {\it projective point\/}
and a plane through the origin is a {\it projective line\/}.
A {\it projective line segment\/} is the portion of
a projective line between two projective points, that is, a pair
of opposite planar wedges with vertex at the origin. 

We can embed affine $d$-space into projective space as a 
hyperplane that misses the origin.  There is a unique
line through any point of this hyperplane and the origin,
and hence each point of affine space corresponds to a unique
projective point.  There is, however, one projective hyperplane,
and many projective $k$-flats for $k < d-1$, without corresponding
affine flats; these are the projective $k$-flats parallel to
the affine space. 
We say that these flats are {\it at infinity\/}.

Each projective point $p$ has a {\it dual\/} projective
hyperplane $D(p)$, namely the hyperplane orthogonal to $p$ 
at the origin in $(d+1)$-dimensional affine space.
Similarly a projective $k$-flat dualizes to its orthogonal $(d-k)$-flat. 
Notice that in projective space, unlike in affine space,
there are no exceptional cases: each $k$-flat is the dual of a $(d-k)$-flat.

Now let $X$ be a set of points in $d$-dimensional projective space. 
(From now on we shall just say ``point'', ``line'', etc.\ rather than
``projective point'', ``projective line'',
 when there is no risk of confusion.)
We now propose a key definition: 
a distance between flats with respect to the points in $X$. 
The definition is more intuitive in the dual formulation than
in the primal, but we give both below for completeness. 
Let $D(F)$ denote the flat that is dual to flat $F$
and let $D(X)$ denote the set of hyperplanes that are dual to points of $X$.
A {\it double wedge\/} is the (closed) region between 
two projective hyperplanes.

\begin{defn}
The {\bf crossing distance} between two flats $F$ and $G$
with respect to $X$ is the minimum number of hyperplanes of $D(X)$
intersected by a (closed) projective line segment
with one endpoint on $D(F)$ and the other on $D(G)$. 
In the primal formulation, the crossing distance between $F$ and $G$
is the minimum number of points of $X$ in a double wedge that contains
$F$ in one bounding hyperplane and $G$ in the other.
\end{defn}

We now turn our attention to linear regression
and for ease of understanding, we return temporarily to $d$-dimensional
affine space.
Assume that we designate $k$ dimensions as
{\it independent variables\/} and $d-k$ as {\it dependent variables\/}.
Let $I$ denote the linear subspace spanned by the independent dimensions.
We call a $k$-flat {\it vertical\/} if its projection onto $I$
is not full-dimensional, that is, if its projection is not all of $I$.
For example, let $k=1$ and $d=3$ and think of the
$x$-axis as representing the independent variable;
then any line contained in a vertical
plane (that is, parallel to the $yz$-plane) is vertical.

In projective space, a $k$-flat is vertical if and only if it contains
a point in a particular $(d-k-1)$-flat 
at infinity, which we call the {\it $(d-k-1)$-flat
at vertical infinity\/} and denote by $V_{d-k-1}$.

\begin{defn}
The {\bf regression depth} of a $k$-flat $F$ is 
its crossing distance from $V_{d-k-1}$.
Equivalently, the regression depth of $F$ is the minimum number
of points whose removal makes $F$ into a nonfit, where a {\bf nonfit} 
is a $k$-flat with crossing distance zero from $V_{d-k-1}$. 
\end{defn}

Any $k$-flat at infinity meets $V_{d-k-1}$ and therefore has
depth zero.  Therefore, any method for selecting a $k$-flat
of nonzero regression depth will automatically choose a $k$-flat coming
from the original affine space, rather than one that exists only in the
projective space used for our definitions.

Note that, unlike the case for ordinary least squares, there does not
seem to be any way of solving $k$-flat regression separately for each
dependent variable.  Even for the problem of finding a line in $\R^3$,
combining the solution to two planar regression lines may result
in a nonfit.

\omit{
Note that, although ordinary least squares regression is normally
defined in terms of dependent and independent variables, one can also
define a variant called ``total least squares'' in which the residual
error is just the Euclidean distance of a data point from the regression
plane, so all coordinates are treated equally.  Total least squares
regression is well solved by singular value decomposition, and in fact
the affine subspaces spanned by the eigenvalues ("principal components") 
are the optimal flats for each dimension~$k$.

\begin{open}
Is there a natural analogue of regression depth for $k$-flats
which treats each coordinate equally rather than partitioning the
coordinates into independent and dependent variables?
\end{open}
}

\section{Nontransversal Families}\label{lemsec}

In order to prove that deep $k$-flats exist,
we need some combinatorial lemmas on large subsets of points without a
hyperplane transversal.

\begin{defn}
Let $S$ be a set of points.  Then we say that a hyperplane $H$ {\bf
cuts} $S$ if both of the two open halfspaces bounded by $H$ contain at
least one point of $S$.  We say that a family of sets
is {\bf transversal} if there is a hyperplane that cuts all sets in the
family.
\end{defn}

\begin{lemma}[\cite{YaoYao-STOC-85,Mat-DCG-92}]\label{partition}
Let $d$ be a constant, and assume we are given a set of $n$ points in
$\R^d$ and a parameter $p$.  Then we can partition the points into $p$
subsets, with at most $2n/p$ points in each subset, such that any
hyperplane cuts $o(p)$ of the subsets.
\end{lemma}

\begin{lemma}[\cite{AloKal-DCG-95}]\label{transversal}
Let $p\ge q>d$ be constants.
Then there is a constant $C(p,q,d)$ with the following property:
If $\cal F$ is any family of point sets in $\R^d$, such that
any $p$-tuple of sets in $\cal F$ contains a transversal subfamily of
$q$ sets, then $\cal F$ can be partitioned into $C(p,q,d)$ transversal
subfamilies.
\end{lemma}

\begin{figure}[t]
$$\includegraphics[width=4.5in]{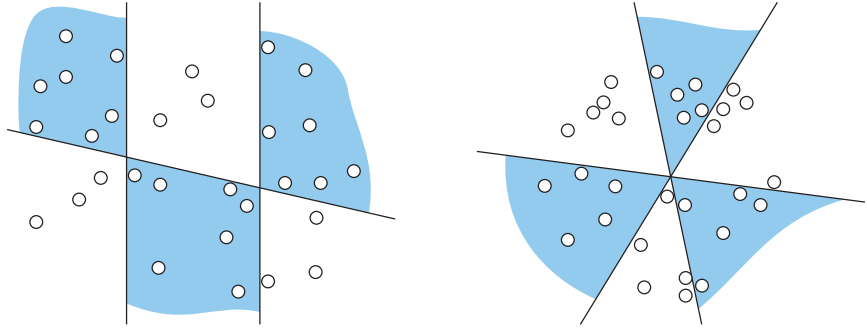}$$
\caption{Construction of three non-transversal sets of $n/6$ points in
$\R^2$: (a) catline, formed by partitioning the point set vertically
into equal thirds, and making a ham sandwich cut of the leftmost and
rightmost $2n/3$ points; (b) subdivision by three coincident lines into
equal sixths.}
\label{sixths}
\end{figure}

\begin{theorem}\label{simplex}
Let $d$ be a constant.
Then there is a constant $P(d)$ with the following property:
For any set $S$ of $n$ points in $\R^d$, we can find a
non-transversal family of
$d+1$ subsets of $S$, such that each subset in the family contains at
least $\lceil n/P(d)\rceil$ points of $S$.
\end{theorem}

\begin{proof}
Choose $p$ to be a multiple of three, sufficiently large that the $o(p)$
bound of Lemma~\ref{partition} is strictly smaller than
$p/(3C(d+1,d+1,d))$, and let $P(d)=2p$.
By Lemma~\ref{partition}, partition $S$ into $p$ subsets of at most
$2n/p$ points, such that any hyperplane cuts few subsets.

Let $\cal F$ be the family consisting of the largest $p/3$ subsets in
the partition.
If the smallest member of $\cal F$ contains $m$ points, then the
total size of all the members of the partition would have to be at most
$(p/3)\cdot 2n/p + (2p/3)\cdot m=2n/3+2pm/3$, but this total size is just
$n$, so $m\ge n/(2p)$ and each member of $\cal F$ contains at least
$n/P(d)$ points.

If each $(d+1)$-tuple of sets in $\cal F$ were transversal,
we could apply Lemma~\ref{transversal} and partition $\cal F$
into $C(p,q,d)$ transversal subfamilies, one of which would have to
contain at least $|{\cal F}|/C(p,q,d)=p/(3C(d+1,d+1,d))$ subsets.
But this violates the $o(p)$ bound of Lemma~\ref{partition},
so $\cal F$ must contain a non-transversal $(d+1)$-tuple.
This tuple fulfills the conditions of the statement of the lemma.
\end{proof}

Clearly, $P(1)=2$ since the median partitions any set of points on a
line into two nontransversal subsets. Figure~\ref{sixths} depicts two
different constructions showing that $P(2)\le 6$.
Although the bound of six is tight for these two constructions
(as can be seen by the example of points equally
spaced on a circle) we do not know whether there might be a different
construction that achieves a better bound;
the best lower bound we have found is the following:

\begin{figure}[t]
$$\includegraphics[height=2in]{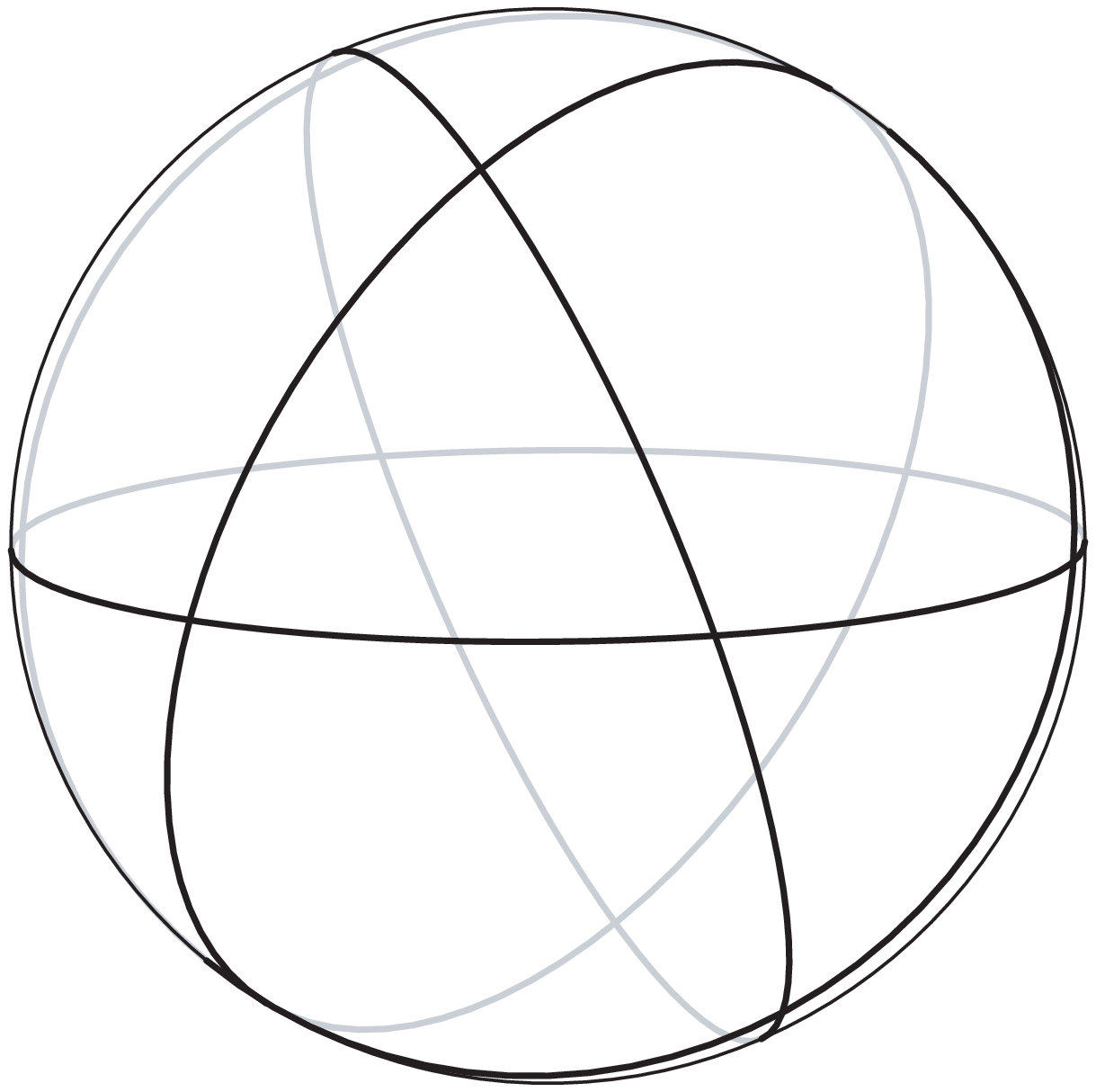}
\qquad
\includegraphics[height=2in]{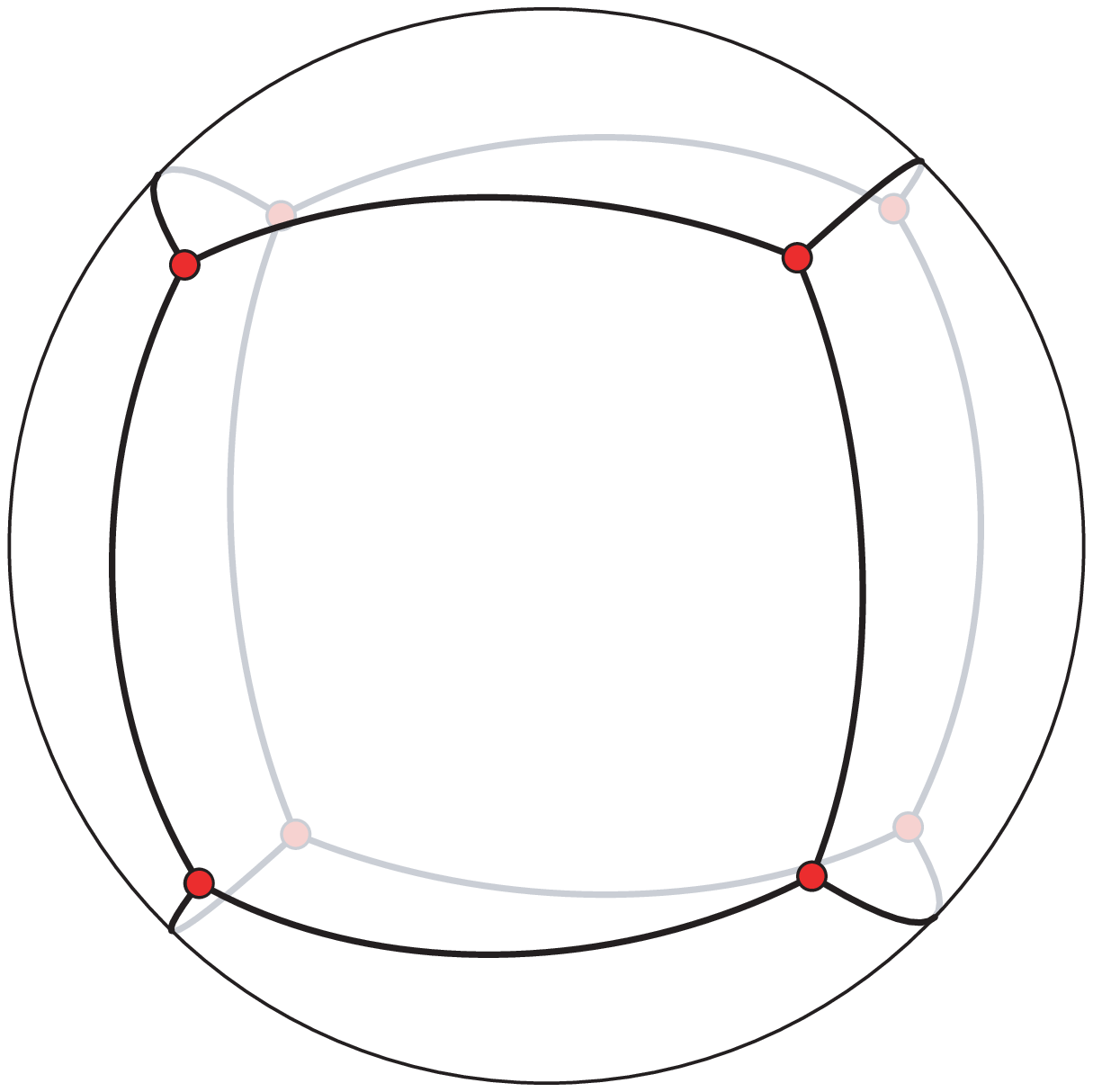}$$
\caption{(a) Four great circles subdivide a sphere in the pattern of a
cuboctahedron. (b) Four pairs of opposite points, connected by great
circle arcs in the pattern of a cuboid.}
\label{spheres}
\end{figure}

\begin{theorem}
$\displaystyle P(2) \ge \frac{\pi}{2\sin^{-1}\frac13} \approx 4.622$.
\end{theorem}

\begin{proof}
We form a distribution on the plane by centrally projecting the uniform
distribution on a sphere.  We show that any nontransversal triple for
this distribution must have a set with measure at most
$1/4.622$ times the total measure.  The same bound then holds in the
limit for discrete point sets formed by taking $\epsilon$-approximations
of this distribution.

Let $S_i$, $i\in\{1,2,3\}$ denote the three nontransversal subsets of
the plane maximizing the minimum measure of any $S_i$.  Without loss of
generality, each $S_i$ is convex.  Consider the three lines tangent to
two of the $S_i$, and separating them from the third set (such a line
must exist since the sets are nontransversal).  These lines form an
arrangement with seven (possibly degenerate) faces: a triangle adjacent
on its edges to three three-sided infinite cells, and on its vertices to
three two-sided infinite cells. The sets $S_i$ coincide with the
three-sided infinite cells: any set properly contained in such a cell
could be extended to the whole cell without violating the
nontransversality condition, and if they instead coincided with the
two-sided cells we could shrink the arrangement's central triangle while
increasing the sizes of all three $S_i$. The two arrangements in
Figure~\ref{sixths} can both be viewed as such three-line
arrangements, degenerate in different ways.

Each line in the plane lifts by central projection to a great circle on
the sphere.  Consider the great circles formed by lifting four lines:
the three lines considered above and the line at infinity.  Any
arrangement of four circles on the sphere cuts the sphere in the pattern
of a (possibly degenerate) cuboctahedron (Figure~\ref{spheres}(a)).
The three-sided infinite cells in the plane lift to quadrilateral faces
of this cuboctahedron. Note that the area of a spherical quadrilateral
is the sum of its internal angles, minus $2\pi$.

Form the dual of the arrangement by treating each great circle as an
``equator'' and placing a pair of points at the corresponding two poles.
The geodesics between these points have the pattern of a cuboid
(Figure~\ref{spheres}(b)) such that the length of each geodesic is an
angle complementary to one of the cuboctahedron quadrilaterals' internal
angles.  Thus, the cuboid minimizing the maximum quadrilateral perimeter
corresponds to the cuboctahedron maximizing the minimum quadrilateral
area. But any spherical cuboid has at least one face covering at least
one-sixth of the sphere, and the minimum perimeter for such a
quadrilateral is achieved when the quadrilateral is a square.
Therefore, the regular cube minimizes the maximum perimeter
and the regular cuboctahedron maximizes the minimum area.
The ratio of a regular cuboctahedron's square face area to the area of a
full hemisphere is the value given,
$\frac{\pi}{2\sin^{-1}{1/3}} \approx 4.622$.
\end{proof}

We also do not know tight bounds on $P(d)$ for $d\ge 3$.
The proof of Theorem~\ref{simplex} (using the best known bounds in
Lemma~\ref{partition}~\cite{Mat-DCG-92}) leads to upper bounds of the form
$O(C(d+1,d+1,d)^{-d})$. We may be able to improve this somewhat, to
$O(C(d+1,d,d-1)^{1-d})$, by a more complicated
construction: Project the points arbitrarily onto a $(d-1)$-dimensional
subspace, find a partition in the subspace, and use
Lemma~\ref{transversal} to find a family of $d+1$ subsets
such that no subfamily of $d$ subsets has a transversal.
As in the catline construction~\cite{HubRou-JMA-98},
group these subsets into $d$ pairs, and form a ham sandwich cut in
$\R^d$ of these pairs, in such a way that this cut partitions each subset
of the family in the same proportion $a:b$, and such that the
half-subsets of size $a$ are above or below the ham sandwich cut
accordingly as the members of the family of $d+1$ subsets are on one or
the other side of a Radon partition of those subsets in $\R^{d-1}$.
Without loss of generality, $a>b$; then choose the $d+1$ subsets
required by Lemma~\ref{simplex} to be the ones of size $a$.

\begin{open}
Prove tighter bounds on $P(d)$ for $d\ge 2$.
\end{open}

\section{Deep $k$-Flats}\label{ksec}

It is previously known that deep $k$-flats exist
for $k=0$~\cite{Rad-JMLS-46} and
$k=d-1$~\cite{AmeBerEpp-DCG-?,SteWen-CCCG-98}.  In this section we show
that such flats exist for all other values of $k$.

We first need one more result, a common generalization of centerpoints
and the ham sandwich theorem:

\begin{lemma}[The Center Transversal
Theorem~\cite{ZivVre-BLMS-90,Dol-MZ-92}]
\label{toothpick}
Let $k+1$ point sets be given in $\R^d$, each containing at least $m$
points, where $0\le k<d$. Then there exists a $k$-flat $F$ such that any
closed halfspace containing $F$ contains at least $\lceil m/(d-k+1)\rceil$
points from each set.
\end{lemma}

The weaker bound $\lceil m/(d+1)\rceil$ can be proven
simply by choosing a flat through the centerpoint of each subset.

\begin{theorem}\label{main}
Let $d$ and $0\le k<d$ be constants.
Then there is a constant $R(d,k)$ such that
for any set of $n$ points with $k$ independent and $d-k$ dependent
degrees of freedom, there exists
a $k$-flat of regression depth at least $\lceil n/R(d,k) \rceil$.
\end{theorem}

\begin{proof}
Project the point set 
onto the subspace spanned by the $k$
independent directions, in such a way that the inverse image of each
point in the projection is a $(d-k)$-flat containing $V_{d-k-1}$.
By Theorem~\ref{simplex},
we can find a family of
$k+1$ subsets of the data points, each with $n/P(k)$ points,
such that the $k$-dimensional projection of this family has no
transversal.  We then let $F$ be the $k$-flat determined by applying
Lemma~\ref{toothpick} to this family of subsets.

Then consider any double wedge bounded by a hyperplane containing $F$ and
a hyperplane containing $V_{d-k-1}$.  The vertical boundary of this
double wedge projects to a hyperplane in $\R^k$, so it must miss one
of the $k+1$ subsets in the family.  Within this missed subset the double
wedge appears to be simply a halfspace through $F$.  By
Lemma~\ref{toothpick}, the double wedge must therefore contain at least
$n/((d-k+1)P(k))$ points.  Thus if let
$R(d,k)=(d-k+1)P(k)$ the theorem is satisfied.
\end{proof}

For $k=0$ or $k=d-1$ we know that $R(d,k)=d+1$
\cite{AmeBerEpp-DCG-?}.
However exact values are not known for
intermediate values of $k$.
\begin{open}
Prove tighter bounds on $R(d,k)$ for $1\le k\le d-2$.
\end{open}

The following conjecture would follow from the assumption that $R(d,k)$
is a linear function of $d$ for fixed $k$ (as the $O(d)$ bound of
Theorem~\ref{main} makes plausible), since $R(k,k)=1$ and $R(k+1,k)=k+2$.
It also matches the known results $R(d,0)=R(d,d-1)=d+1$ and
the bounds $R(d,1)\le 2d-1$ and $R(3,1)=5$ of the following section.

\begin{conjecture}\label{exact}
$R(d,k)=(k+1)(d-k)+1$.
\end{conjecture}

\section{Tighter Bounds for Lines}\label{linesec}

The proof of Theorem~\ref{main} shows that
$R(d,1)\le 2d$; this can
be slightly improved using a technique of overlapping sets borrowed from
the catline construction.

\begin{theorem}\label{upperline}
$R(d,1)\le 2d-1$.
\end{theorem}

\begin{proof}
The proof of Theorem~\ref{main} can be viewed as projecting the points
onto a horizontal line, dividing the line into two rays at the median of
the points, and applying the center transversal theorem to the
two sets of $n/2$ points contained in each ray.
Instead, we project the points onto the horizontal line as before,
but partition this line into three pieces: two rays
containing $(d-1)n/(2d-1)$ points each, and a line segment in the
middle containing the remaining $n/(2d-1)$ points.
We then apply the center transversal theorem to two sets $S_1$ and
$S_2$ of
$dn/(2d-1)$ points each, formed by the points having a projection in the
union of the middle segment and one of the two rays.  This theorem finds
a line such that no halfspace containing it has fewer than $n/(2d-1)$
points in either of the sets $S_i$. We claim that this line has
regression depth at least
$n/(2d-1)$.

To prove this, consider any double wedge in which one hyperplane
boundary contains the regression line, and the other hyperplane boundary
is vertical.  The vertical hyperplane then intersects the horizontal
projection line in a single point. If this intersection point is in one
of the two rays, then the vertical hyperplane misses the set $S_i$ formed
by the other ray and the middle segment.  In this case, the double wedge
contains the same subset of $S_i$ as a halfspace bounded by the
double wedge's other bounding plane, and so contains at least $n/(2d-1)$
points of $S_i$

In the remaining case, the vertical boundary of the double wedge
intersects the horizontal projection line in its middle segment.
Within each of the two sets to which we applied the center transversal
theorem, the double wedge differs from a halfspace (bounded by the
same nonvertical plane) only within the middle set.  But any hyperplane
bounds two different halfspaces, and the halfspace approximating the
double wedge in $S_1$ is opposite the halfspace approximating the double
wedge in $S_2$. Therefore, if we
let $X_i$ denote the set of points in the halfspace but not in the
double wedge, then $X_1$ and $X_2$ are disjoint subsets of the middle
$n/(2d-1)$ points.  The number of points in the double wedge within one
set $S_i$ must be at least $n/(2d-1) - |X_i|$,
so the total number of points in the double wedge
is at least $2n/(2d-1)-|X_1\cup X_2|\le 2n/(2d-1)-n/(2d-1)=n/(2d-1)$.

Thus in all cases a double wedge bounded by a hyperplane through the
regression line and by a vertical hyperplane contains at least $n/(2d-1)$
points, showing that the line has depth at least $n/(2d-1)$.
\end{proof}

\begin{figure}[t]
$$\includegraphics[width=4.5in]{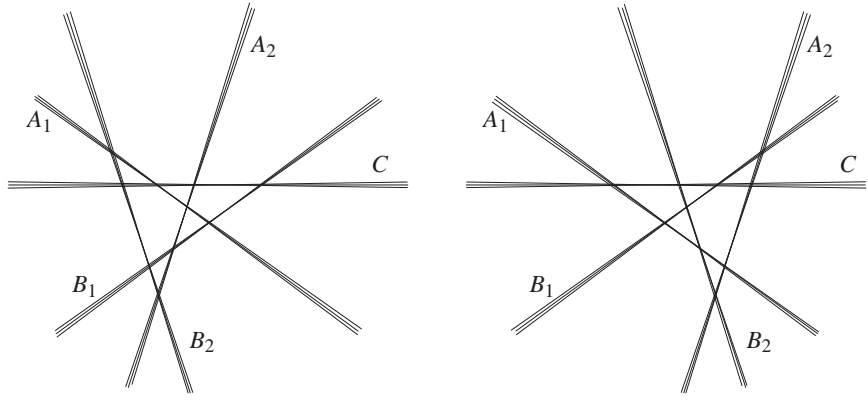}$$
\caption{Three-dimensional plane arrangement showing $R(3,1)\ge 5$:
(a) cross-section for $x=1$; (b) cross-section for $x={-1}$.}
\label{fifths}
\end{figure}

As evidence that this $2d-1$ bound may be tight, we present a matching
lower bound for $d=3$.

\begin{theorem}
$R(3,1)=5$.
\end{theorem}

\begin{proof}
We have already proven that $R(3,1)\le 5$, so we need only show that
$R(3,1)\ge 5$. We work in the dual space, and construct an arrangement of
$n$ planes in
$\R^3$, for $n$ any multiple of 5,
such that any line has depth at most $n/5$.

Our arrangement consists of five groups of nearly parallel and
closely spaced planes, which we label $A_1$, $A_2$, $B_1$, $B_2$, and $C$.
Rather than describe the whole arrangement, we describe the line arrangements
in the planar cross-sections at $x=1$ and
$x={-1}$.
Recall that the depth of a line in the
three-dimensional arrangement is the minimum number of planes crossed by
any vertical ray starting on the line. Limiting attention to rays
contained in the two cross-sections (and hence to the planar depth of the
two points where the given line intersects these cross-sections)
gives an upper bound on the depth of the line, and so a lower bound on
$R(3,1)$.

In the first cross-section, we place the groups of lines as shown in
Figure~\ref{fifths}(a), with the region where $A_1$ and $A_2$ cross
contained inside the triangle formed by the other three groups. 
Moreover, $A_1$ and $A_2$ do not cross side $B_2$ of the
triangle, instead crossing group $B_2$ at points outside
the triangle.  The points where members of $A_1$ intersect each other
are positioned on segment $CA_1$ -- $A_1A_2$.
Similarly, the crossings within $A_2$ are situated
on segment $A_1A_2$ -- $A_2B_1$.
The crossings within $B_1$, $B_2$, and $C$ are
situated along the corresponding sides of the triangle formed by these
three groups.

In the cross-section formed as described above, points from most cells in
the arrangement can reach infinity while crossing only one group, and so
have depth at most $n/5$.  It is only within the segments
$CA_1$ -- $A_1A_2$ and  $A_1A_2$ -- $A_2B_1$
that a point can have higher depth.
The arrangement is qualitatively similar for nearby cross-sections
$x=1\pm\epsilon$.
Therefore, any deep line in $\R^3$ must be either nearly parallel to $A_1$
and not near any $B_i$, or nearly parallel to $A_2$ and not near $B_2$.

In the second cross-section (Figure~\ref{fifths}(b)), the groups $A_i$
and $B_i$ reverse roles: the point where $B_1$ and $B_2$ cross is
contained in the triangle determined by the other three groups, and the
other details of the arrangement are situated in a corresponding manner.
Therefore, any deep line would have to be either nearly parallel to
$B_1$ and not near any $A_i$, or nearly parallel to $B_2$ and not
near $A_2$.

There is no difficulty forming
the two cross-sections described above from a single plane arrangement,
since (as shown in
Figures 3(a) and (b)) the slopes of the lines within each group can
remain the same in each cross-section.
But the requirements imposed on a deep line by these two cross-sections
are contradictory, therefore no line can have depth greater than $n/5$
in this arrangement.
\end{proof}

We believe that a similar proof can be used to prove a more general
$2d-1$ lower bound on $R(d,1)$ in any dimension, matching the upper
bound in Theorem~\ref{upperline}:
form an arrangement with hyperplane groups $A_i$, $B_i$, and $C$,
so that in one cross-section the $A_i$ meet in a vertex contained in a
simplex formed by the other groups, and in the other cross-section the
groups $A_i$ and $B_i$ exchange roles.  However we have not worked out
the details of where to place the intersections within groups, how to
choose hyperplane angles such that the inner groups miss a face of the
outer simplex in both cross-sections, or which cells of the resulting
arrangements can have high depth.

\section{Generalizations of Tverberg's Theorem}\label{tversec}

A {\em Tverberg partition} of a set of point sites is a partition of the
sites into subsets, the convex hulls of which all have a common
intersection. The {\em Tverberg depth} of a point $t$ is the maximum
cardinality of any Tverberg partition for which the common intersection
contains $t$. Note that the Tverberg depth is a lower bound on the
location depth. Tverberg's theorem \cite{Tve-JLMS-66, Tve-BAMS-81} is
that there always exists a point with Tverberg depth $\lceil
n/(d+1)\rceil$ (a {\em Tverberg point}); this result generalizes both the
existence of center points (since any Tverberg point must be a center
point) and Radon's theorem \cite{Rad-MA-21} that any $d+2$ points have a
Tverberg partition into two subsets.

Another way of expressing Tverberg's theorem is that for any point set we
can find both a partition into $\lceil n/(d+1)\rceil$ subsets, and a point
$t$, such that $t$ has nonzero depth in each subset of the partition. 
Stated this way, there is a natural generalization to higher dimensional
flats:

\begin{theorem}
Let $d$ and $0\le k<d$ be constants.
Then there is a constant $T(d,k)$
such that
for any set of $n$ points with $k$ independent and $d-k$ dependent
degrees of freedom, there exists
a $k$-flat $F$ and a partition of the points into
$\lceil n/T(d,k)\rceil$ subsets,
such that $F$ has nonzero regression depth in each subset.
\end{theorem}

\begin{proof}
As in the proof of Theorem~\ref{main}, we
project the point set 
onto the subspace spanned by the $k$
independent directions, in such a way that the inverse image of each
point in the projection is a $(d-k)$-flat containing $V_{d-k-1}$.
By Theorem~\ref{simplex},
we can find a family of
$k+1$ subsets $S_i$, each with $n/P(k)$ points,
such that the $k$-dimensional projection of this family has no
transversal.  We then find a Tverberg point $t_i$ and a Tverberg partition
of each set $S_i$ into subsets $T_{i,j}$, for $1\le j\le
\lceil n/(P(k)(d+1))\rceil$. We let $F$ be the
$k$-flat spanning these $k+1$ Tverberg points. We form each set 
$T_j$ in our
Tverberg partition as the union $\cup_i T_{i,j}$.  Some points of $S$ may
not belong to any set $T_{i,j}$, in which case they can be assigned
arbitrarily to some set $T_i$ in the partition.

Then consider any double wedge bounded by a hyperplane containing $F$ and
a hyperplane containing $V_{d-k-1}$.  The vertical boundary of this
double wedge projects to a hyperplane in $\R^k$, so it must miss one
of the $k+1$ subsets $S_i$.  Within $S_i$ the double
wedge appears to be simply a halfspace through $t_i$.
It therefore contains at least one point of each set $T_{i,j}$ and
a fortiori at least one point of each set $T_j$.
Thus if let
$R(d,k)=(d+1)P(k)$ the theorem is satisfied.
\end{proof}

We know that $T(d,0)=d+1$ by Tverberg's theorem, and the
catline construction~\cite{HubRou-JMA-98}
shows that $T(2,1)=3$.
However even in the case $k=d-1$ we do not know a tight
bound; Rousseeuw and Hubert~\cite{RouHub-JASA-99, RouHub-DCG-99}
conjectured that $T(d,d-1)=d+1$ but
the best known bounds from our previous paper
\cite{AmeBerEpp-DCG-?} are $T(d,d-1)\le d(d+1)$
and $T(3,2)\le 6$.  

\begin{open}
Prove tighter bounds on $T(d,k)$ for $1\le k\le d-1$.
\end{open}

\section{Connection Between $k$-Flats and $(d-k-1)$-Flats}\label{revsec}

There is a natural relation between finding a deep $k$-flat and finding
a deep $(d-k-1)$-flat: in both cases one wants to find a $k$-flat and a
$(d-k-1)$-flat that are far apart from each other, and the problems
only differ in which of the two flats is fixed at vertical infinity,
and which is to be found.

In our previous paper \cite{AmeBerEpp-DCG-?} we exploited this
connection in the following way, to show that $R(d,d-1)=d+1$.
A centerpoint (corresponding to the value of $R(d,0)$) is just
a point far from a given ``hyperplane at infinity''; in projective
$d$-space, this hyperplane can be chosen arbitrarily, resulting in
different centerpoint locations. We first found an appropriate way to
replace the input point set by a smooth measure, and modify the
definition of a centerpoint, in such a way that we could show that the
modified centerpoint location varied continuously and equivariantly, as a
function of the position of the hyperplane at infinity.
In oriented projective space, the set of hyperplane locations and the
set of centerpoint locations are both topological $d$-spheres,
so we could have applied the Borsuk-Ulam theorem (in the form that every
continuous equivariant function from the sphere to itself is surjective)
to find a hyperplane in the inverse image of the point at vertical
infinity; this hyperplane is the desired deep regression plane. Our actual
proof used the Brouwer fixed point theorem in a similar way, avoiding the
need to use the equivariance property.

Conjecture~\ref{exact} implies in more generality that
$R(d,k)=R(d,d-k-1)$,
and one would naturally hope for a similar proof of this equality. There
are two obstacles to such a hope: First, we do not know how to modify the
definition of a deep
$k$-flat in such a way as to choose a unique flat which varies
continuously as a function of the location of the
$(d-k-1)$-flat at infinity.  A similar lack of a continuous version of
Tverberg's theorem blocked our attempts to prove that $T(d,d-1)=d+1$.
However, some of our constructions (for instance the bound
$2(d+1)$ on $R(d,1)$ formed by vertically bisecting the points and
choosing a line through the centerpoints of each half) can be made
continuous using ideas from our previous paper.  Second, and more
importantly, the space $F_k^d$ of oriented $k$-flats does not form a
topological sphere, and there can be continuous equivariant
non-surjective functions from this space to itself.  Nevertheless there
might be a way of using generalizations of the Borsuk-Ulam
theorem~\cite{HusLasMag-JME-90} or a modification of our Brouwer fixed
point argument to show that the deep $k$-flat function must be
surjective, perhaps using the additional property that a deep $k$-flat
cannot be incident to $V_{d-k-1}$.

\begin{open}
Can there exist a continuous non-surjective $\Z_2$-equivariant map
$c$ from $F_{d-k-1}^d$ to $F_k^d$ such that any $(d-k-1)$-flat $V$
and its image $c(V)$ are never incident?
\end{open}

\begin{open}
Does $R(d,k)=R(d,d-k-1)$ for $1\le k\le d-2$?
\end{open}

\begin{open}
Does $T(d,k)=T(d,d-k-1)$ for $0\le k\le d-1$?
\end{open}

\section{Algorithmic Implications}\label{algsec}

We now show how to use our proof that deep flats exist
as part of an algorithm for finding an approximate deepest flat.
We begin with an inefficient exact algorithm.

\begin{theorem}\label{exact-alg}
Let $d$ and $k$ be constants.  Then we can find the deepest
$k$-flat for a collection of $n$ points in $\R^d$,
in time $n^{O(1)}$.
\end{theorem}

\begin{proof}
Let $A$ be the arrangement of hyperplanes dual to the $n$ given points.
The distance from points in $\R^d$ to $V_{d-k-1}$
is constant within each cell of $A$, and all such distances
can be found in time $O(n^d)$ by applying a breadth first search
procedure to the arrangement.  The depth of a $k$-flat $F$ is just the
minimum depth of any cell of $A$ pierced by $F$.
Any two flats that pierce the same set of cells of $A$ have the same
depth.

The space of $k$-flats forms a $(k+1)(d-k)$-dimensional algebraic set
$F_k^d$, in which the flats touching any $(d-k-1)$-dimensional cell of $A$
form a subset of codimension one.  The arrangement of these
$O(n^d)$ subsets partitions $F_k^d$ into
$O(n^{d(k+1)(d-k)+\epsilon})$ cells, corresponding to collections of
flats that all pierce the same set of cells.
We can construct this arrangement,
and walk from cell to cell maintaining a priority queue of the depths of
the cells in $A$ pierced by the flats in the current cell,
in time $O(n^{d(k+1)(d-k)+\epsilon})$.
\end{proof}

We now use standard geometric sampling techniques to combine this exact
algorithm with our lower bound on depth, resulting in an
asymptotically efficient approximation algorithm.

\begin{theorem}\label{approx-alg}
Let $d$, $k$, and $\delta>0$ be constants.  Then we can find the a
$k$-flat with depth within a $(1-\delta)$ factor of the maximum,
for a collection of $n$ points in $\R^d$,
in time $O(n)$.
\end{theorem}

\begin{proof}
We first construct an $\epsilon$-approximation $S$ of the points, 
for the range space consisting of double wedges with one vertical
boundary, where $\epsilon=\delta/(2R(d,k))$.  Then if a flat $F$ has depth
$D$ with respect to $S$,
$Dn/|S|$ is within an additive $\epsilon n$ term of the true
depth of $F$ with respect to the original point set.
$S$ can be found with
$|S|=O(\epsilon^{-2}\log\epsilon^{-1})$, in time $O(\epsilon)$,
using standard geometric sampling algorithms.
We then let $F$ be the deepest flat for $S$.

Suppose the optimal flat $F^*$ for the original point set
has depth $cn$.  Then the depth of $F^*$ in $S$,
and therefore also the depth of $F$ in $S$,
must be at least $(c-\epsilon)|S|$.
Therefore, the depth of $F$ in the original point set
must be at least $(c-2\epsilon)n$.
Since $c\ge 1/R(d,k)$,
$(c-2\epsilon)n\ge (1-\delta)cn$.
\end{proof}

Although our approximation algorithm takes only linear time,
it is likely not practical due to its high constant factors.
However, perhaps similar ideas can form the basis of a more
practical random sampling based algorithm.

\begin{open}
Improve the time bounds on finding an exact deepest $k$-flat.
Is it any easier to find a $k$-flat with depth
at least $n/R(d,k)$, that may not necessarily approximate the deepest
flat?
\end{open}

\subsection*{Acknowledgements}

Work of Eppstein was done in part while visiting Xerox PARC.
The authors would like to thank Peter Rousseeuw for helpful
comments on a draft of this paper.

\bibliographystyle{abuser}
\bibliography{regdepth}

\begin{thebibliography}{10}

\bibitem{AloKal-DCG-95}
N.~Alon and G.~Kalai.
\newblock {Bounding the piercing number}.
\newblock {\em Discrete {\&} Computational Geometry} 13(3--4):245--256, 1995.

\bibitem{AmeBerEpp-DCG-?}
N.~Amenta, M.~Bern, D.~Eppstein, and S.-H. Teng.
\newblock {Regression depth and center points}.
\newblock To appear in {\em Discrete {\&} Computational Geometry},
  cs.CG/9809037.

\bibitem{Dol-MZ-92}
V.~L. Dol'nikov.
\newblock {A generalization of the sandwich theorem}.
\newblock {\em Mat. Zametki} 52(2):27--37, 1992.

\bibitem{HubRou-JMA-98}
M.~Hubert and P.~J. Rousseeuw.
\newblock {The catline for deep regression}.
\newblock {\em J. Multivariate Analysis} 66:270--296, 1998,
  \url{http://win-www.uia.ac.be/u/statis/publicat/catline_abstr.html}.

\bibitem{HusLasMag-JME-90}
S.~Y. Husseini, J.-M. Lasry, and M.~J.~P. Magill.
\newblock {Existence of equilibrium with incomplete markets}.
\newblock {\em J. Mathematical Economics} 19:39--67, 1990.

\bibitem{KreMitRou-SCG-99}
M.~van Kreveld, J.~S.~B. Mitchell, P.~J. Rousseeuw, M.~Sharir, J.~Snoeyink, and
  B.~Speckmann.
\newblock {Efficient algorithms for maximum regression depth}.
\newblock {\em Proc. 15th Symp. Computational Geometry}, pp. 31--40. ACM, June
  1999.

\bibitem{LanSte-SODA-00}
S.~Langerman and W.~Steiger.
\newblock {An $O(n\log n)$ algorithm for the hyperplane median in $\R^2$}.
\newblock {\em Proc. 11th Symp. Discrete Algorithms}. ACM and SIAM, 2000.

\bibitem{Liu-Stat-92}
Z.~J. Liu.
\newblock {Robustness of least distances estimate in multivariate linear
  models}.
\newblock {\em Statistics} 23(2):109--119, 1992.

\bibitem{Mat-DCG-92}
J.~Matou{\v{s}}ek.
\newblock {Efficient partition trees}.
\newblock {\em Discrete {\&} Computational Geometry} 8(3):315--334, 1992.

\bibitem{Rad-JMLS-46}
R.~Rado.
\newblock {A theorem on general measure}.
\newblock {\em J. London Math. Soc.} 21:291--300, 1946.

\bibitem{Rad-MA-21}
J.~Radon.
\newblock {Mengen konvexer K{\"o}rper, die einen gemeinsamen Punkt Enthalten}.
\newblock {\em Math. Annalen} 83:113--115, 1921.

\bibitem{Rou-JASA-84}
P.~J. Rousseeuw.
\newblock {Least median of squares regression}.
\newblock {\em J. Amer. Statistical Assoc.} 79:871--880, 1984.

\bibitem{RouHub-DCG-99}
P.~J. Rousseeuw and M.~Hubert.
\newblock {Depth in an arrangement of hyperplanes}.
\newblock {\em Discrete {\&} Computational Geometry} 22:167--176, 1999,
  \url{http://win-www.uia.ac.be/u/statis/publicat/arrang_abstr.html}.

\bibitem{RouHub-JASA-99}
P.~J. Rousseeuw and M.~Hubert.
\newblock {Regression depth}.
\newblock {\em J. Amer. Statistical Assoc.} 94(446):388--402, June 1999,
  \url{http://win-www.uia.ac.be/u/statis/publicat/rdepth_abstr.html}.

\bibitem{RouStr-SC-98}
P.~J. Rousseeuw and A.~Struyf.
\newblock {Computing location depth and regression depth in higher dimensions}.
\newblock {\em Statistics and Computing} 8(3):193--203, August 1998,
  \url{http://win-www.uia.ac.be/u/statis/publicat/compdepth_abstr.html}.

\bibitem{SteWen-CCCG-98}
W.~Steiger and R.~Wenger.
\newblock {Hyperplane depth and nested simplices}.
\newblock {\em Proc. 10th Canad. Conf. Computational Geometry}. McGill Univ.,
  1998,
  \url{http://cgm.cs.mcgill.ca/cccg98/proceedings/cccg98-steiger-hyperplane.ps%
.gz}.

\bibitem{Tve-JLMS-66}
H.~Tverberg.
\newblock {A generalization of Radon's theorem}.
\newblock {\em J. London Math. Soc.} 41:123--128, 1966.

\bibitem{Tve-BAMS-81}
H.~Tverberg.
\newblock {A generalization of Radon's theorem II}.
\newblock {\em Bull. Austral. Math. Soc.} 24(3):321--325, 1981.

\bibitem{YaoYao-STOC-85}
A.~C. Yao and F.~F. Yao.
\newblock {A general approach to $d$-dimensional geometric queries (extended
  abstract)}.
\newblock {\em Proc. 17th Symp. Theory of Computing}, pp. 163--168. ACM, May
  1985.

\bibitem{ZivVre-BLMS-90}
R.~T. {\v{Z}}ivaljevi{\'c} and S.~Vre{\'{c}}ica.
\newblock {An extension of the ham sandwich theorem}.
\newblock {\em Bull. London Math. Soc.} 22(2):183--186, 1990.

\end{thebibliography}
\end{document}